\documentclass{article}
\reversemarginpar
\newcommand{\bq}{\begin{quotation}}
\newcommand{\eq}{\end{quotation}}
\newcommand{\bi}{\begin{itemize}}
\newcommand{\ei}{\end{itemize}}
\newcommand{\bc}{\begin{center}}
\newcommand{\ec}{\end{center}}
\newcommand{\noi}{\noindent}
\newcommand{\bt}{\begin{tabular}}
\newcommand{\et}{\end{tabular}}
\newcommand{\bee}{\begin{eqnarray}}
\newcommand{\eee}{\end{eqnarray}}
\newcommand{\ba}{\begin{array}}
\newcommand{\ea}{\end{array}}
\newcommand{\be}{\begin{equation}}
\newcommand{\ee}{\end{equation}}
\newcommand{\ssigma}{\mbox{\boldmath $\sigma$}}

\newcommand{\nnabla}{\mbox{\boldmath $\nabla$}}
\newcommand{\oOmega}{\mbox{\boldmath $\Omega$}}

\newcommand{\br}{\mbox{\boldmath $r$}}
\newcommand{\bw}{\mbox{\boldmath $w$}}
\newcommand{\bv}{\mbox{\boldmath $v$}}
\newcommand{\bU}{\mbox{\boldmath $U$}}
\newcommand{\bcF}{\mbox{\bf f}}
\newcommand{\bR}{\mbox{\boldmath $R$}}
\newcommand{\baa}{\mbox{\boldmath $a$}}
\newcommand{\bb}{\mbox{\boldmath $b$}}
\newcommand{\bone}{\mbox{\boldmath $1$}}
\newcommand{\bP}{\mbox{\boldmath ${\cal P}$}}
\newcommand{\bG}{\mbox{\boldmath ${\cal G}$}}
\newcommand{\bF}{\mbox{\boldmath $F$}}
\newcommand{\bZ}{\mbox{\boldmath $Z$}}

\newcommand{\btan}{\mbox{\boldmath $t$}}
\newcommand{\bn}{\mbox{\boldmath $n$}}
\newcommand{\bT}{\mbox{\boldmath $T$}}
\newcommand{\bC}{\mbox{\boldmath $C$}}
\newcommand{\bE}{\mbox{\boldmath $E$}}
\newcommand{\bV}{\mbox{\boldmath $V$}}

\newcommand{\xb}{\mbox{\boldmath $\zeta$}}
\begin{document}
\bc
\huge 
{\bf 
Hydrodynamic interactions \\
between many spheres}\\ 
\ec
\bc
\Large
{\bf Maria L. Ekiel-Je\.zewska}\\
\large 
PMMH ESPCI, Paris, France\\
(on leave from: IPPT
PAN, Warszawa, Poland)\\
mekiel@ippt.gov.pl

\vspace{0.3cm}
November 20, 1998
\ec

\vspace{0.2cm}
\noi {\large \bf Abstract}

This paper is an introductory guide to many-particle hydrodynamic interactions. 
Basic concepts of the fluid mechanics are assumed to be known. Experience in the Stokes equations is useful but not necessary. 
The study is estimated to fit  
five sessions about three hours each.

\vspace{1cm}
\hfill{\footnotesize \it \parbox{7.5cm}{{\em Auguste Rodin}:  ``Nothing else 
that I have done satisfies me as much, because nothing else cost me so much 
effort...''\cite{Rodf}}}

{\small 
\tableofcontents}

\section{Introduction}
\hfill{\footnotesize \it \parbox{7.5cm}{{\em Auguste Rodin}:  ``...human thought is limited by comparison with what nature transmits directly to us and imposes on us. All that is nesessary is to follow the model; character results from its unity...''\cite{Roda}}}

\vspace{0.5cm}
The text consists of three parts: a brief formulation of the problem, work sheets for own studies aiming to develop the basic concepts and a short concluding overview to indicate how these concepts are useful in construction of the theory modelling quantitatively many-particle systems  \cite{F76a}-\cite{C}.  

The structure of the curriculum emerged from the method 
of effective learning by inquiry \cite{moduly}, redesigned and extended 
according to our previous experience, to specific needs of the subject and to feedback received from participants of a course guided by a preliminary version of the worksheets presented here. 
\cite{moduly}, a non-standard textbook for teachers, called ``a set of laboratory-based modules'' by the authors, provides a method to develop habbits of effective learning, 
based on active inquiry, application of scientific reasoning and cooperative work in small groups. 
Connection with reality 
is essential in this learning pattern, since it provides the motivation leading to a personal engagement (students start from their own obsevations) and it serves as the natural objective 
evaluation of own understanding (students make experiments, which verify predictions of physical models which they have just constructed). A discrepency between own predictions and reality becomes a driving force to learn. 
An extension of this approach to study a theoretical 
cirriculum has been neither straightforward nor simplistic. Actually, 
it required a general analysis of creative learning principles.

The course has been guided by a continuous struggle to 
take care of integrity of the learning process, namely to make its mission 
and its vision clear, its specific goals apparent, 
its structure simple and evolving to fit students' needs and capacities. 
Therefore we have started with a very specific formulation of the goal. We have tried not only to identify and to keep the right sequence of steps building on each other, but also to make this sequence apparent in advance. We have shown how to make various side connections, giving a chance to see possible generalizations and applications, to establish a relation with own experience, and to recognize 
an own direction of further studies.  
The nesessary attitude of the instructor reaching out for integrity was to be first of all a student, challenged to develop a new deeper insight into learning/teaching techniques 
as well as into physical and mathematical aspects of the hydrodynamic interactions, to make the own learning evident, and to allow other learners for 
influencing what and how they learn.  
This curriculum would be never developed without its practical application.

These principles have led to the structure of the learning process presented here. 
The problem and the goal were formulated specifically in the announcement about the course (see Appendix) sent to scientists and students working on problems related to hydrodynamic interactions. This information was important to decide for participation in a non-standard activity. 
Originally the curriculum had been designed to be based only on active group 
work rather than passive listening to lectures. However, the participating scientists 
demanded to be conscious not only of a direction and goals, but also of a perspective of their studies, important in choosing what to investigate 
further. 
Therefore the structure has been modified. 
Work sheets (Sec.~\ref{s2}-\ref{sn42}) served as a guide in own studies 
carried out in small groups of 2-4 people during 
4 sessions about 2.5 hours each. 
An overview concluding 
lecture (Sec.~\ref{sn51}-\ref{sn53}) was added at the end of the course as a closure and as an application of the participants' own inquires. A similar pattern of education had been earlier developed and tested in \cite{GMW}. 


\part{Formulation of the problem}
How to determine the behavior of N spheres in low Reynolds number 
incompressible fluid flow (N up to several hundred)?

We will concentrate on the following 'friction problem':
\\
If translational and rotational velocities of the spheres are given, as well as 
an ambient fluid flow in which they have been immersed, then what are the forces and torques they exert on the fluid?
\bq \footnotesize
This approach can be afterwards adjusted to solve also the twin 'mobility problem': \\
If an ambient fluid flow and external forces and torques acting on the spheres  
are given, then what are their translational and rotational velocities? 
\cite{F88}
\eq

Our goal is to inquire the basic structure and tools of the technique developed 
in \cite{F76a}--\cite{C}.

\part{Developing basic concepts: work sheets}
\hfill{\footnotesize \it \parbox{7.5cm}{{\em Auguste Rodin}:  ``I forced myself to express in each swelling of the torso or of the limbs the efflorescence of a muscle or of a bone which lay deep beneth the skin. And so the truth of my figures, instead of being merely superficial, seems to blossom from within to outside, like life itself.''\cite{Roda}}}

\vspace{0.5cm}
Following \cite{moduly}, in this part we used different type styles to distinguish between a text guiding independent work (written like this sentence), general informations (slanted) and additional remarkes (small letters).

\section{Principles of work}\label{s2}

\vspace{0.5cm}
{\sl The idea is to make the whole problem a subject of your own active inquiry,  
carried out and discussed in small groups of 2-4 persons, on the basis of work sheets written specially for you. 

Questions 
raised during our sessions will help to identify separate steps to be made, building subsequently on each other. 
Each step consists of  the problems  (formulated as a separate subsection of the work sheets) to be solved by you. 
You may find it useful to keep 
a written record of your work.
 
The end of each subsection is a point  
to conclude -- first to share your reasoning with each other, and next to discuss  
your results with me, giving me your comments and questions. 
To keep track of time we will indicate in a 'calendar of progress' when your group 
has finished each subsection. \\
If you have a problem blocking your progress, and any of your group cannot solve 
it, please ask me for help.

Our interests and background vary. Therefore it is reasonable to divide 
into groups of a similar attitude. The content of work sheets is the same for 
all members of a group, but it can be different for different groups, according to your specific needs.  
If you find it useful, don't hesitate to change a group and/or to demand 
for a curriculum related to 
your own questions on the problem.  


There is a collection of references quoted in the instruction. They are to be read only to such an extend which you find relevant and useful to solve the problems posed in the work sheets and to answer your own questions.}

\section{Simplifying: analogy between Stokesian hydrodynamics and electrostatics}\label{s3}
\hfill{\footnotesize \it \parbox{7.5cm}{{\em Auguste Rodin}:  ``The most remote 
antiquity is my habitat. I want to link the past to the present; to return to memory, judge it, and contrive to complete it. Symbols are the guidelines of humanity. They are no lies.
''\cite{Rodf}}}

\vspace{0.5cm}
\subsection{Reasoning by analogy}
Read an introduction from \cite{moduly}, p.~90. 

{\sl Electrostatics is simpler than Stokesian hydrodynamics. Therefore developing the analogy and showing 'how its corresponding parts are alike' helps to understand the basic concepts and processes of the complex technique we are going to study. To put the emphasis on foundations in this section we assume that there is no external ambient flow, which will be added to the system in Sec.~\ref{s4}.}

\subsection{Units}

{\sl To allow for an easy comparison with \cite{Kim} and \cite{Jackson}, our 
reference textbooks, we will use SI units in hydrodynamics and CGS units in electrostatics, i.e. we assume that k=1.
}

How the unit of charge (so-called statcoulomb) is defined in CGS system?\\
How does this unit relate to centimeter, second and gram?\\
Calculate how many coulombs it is.

\noi {\sl Reference:  \cite{Jackson}, Appendix 4}

\subsection{Basic equation}\label{fou}
{\sl The basic equation of both electrostatics and hydrodynamics can be written in a general form as:
\be
L_0 \Psi = s\label{ba}
\ee
where $\Psi$ is a physical field to be found, $L_0$ is a differential operator,
and s is a known source distribution. In electrostatics $\Psi$ is the scalar 
potential field $\Phi$ and $s$ is the charge density $\rho$}.

Specify the operator $L_0$ in electrostatics. Use CGS units. 
  
Specify the meaning of an unknown field $\Psi$, a differential operator $L_0$ and a given source $s$ in Stokesian hydrodynamics. What are the similarities and the differences in comparison to electrostatics?

Compare your analogy with analogies developed by other groups.

{\sl \noi The eq. (\ref{ba}) needs to be supplemented by boundary conditions.}

Specify what do you understand as the fluid boundries in the friction problem.

{\sl \noi References: \cite{Jackson}, Sec.~1.7, \cite{Kim}, Sec.~1.2.3.}

\subsection{Uniqueness theorem}\label{uniq}
\subsubsection*{Guiding question} 
{\sl In electrostatics the solution to the Poisson equation is determined uniquely by specifying on the boundary:\\
-the normal component of the electrostatic field $\bE$   
(so-called Neumann condition) or \\
-the potential $\Phi$ (so-called Dirichlet condition).}

Predict what is a 
hydrodynamic analogue of this theorem.

\subsubsection{Electrostatics}
$\,$\indent Use the Green's identity (\cite{Jackson}, Sec.~1.8): 
\be
\int_V (\phi \nnabla^2 \psi + \nnabla \phi \cdot \nnabla \psi)d^3 \br = \oint_S \phi \nnabla \psi\cdot \bn dA\label{gid}
\ee
(where S is a surface surrounding V, $\bn$ is the unit vector normal to S) to show 
the uniqueness of solutions to the Poisson equation 
(\ref{ba}) for Dirichlet and Neumann boundary conditions (\cite{Jackson}, Sec.~1.9). Specify what is the meaning of uniqueness in both cases. 

\subsubsection{Stokesian hydrodynamics}\label{s3.2.2}
$\,$\indent Develop a similar proof for the Stokesian hydrodynamics. \\
Use the Gauss theorem for a tensor $K$:
\be
\int_V \partial_i K_{i...l}d^3 \br = \oint_S K_{i...l} n_i dA
\ee
to formulate a useful generalization of the Green's identity (\ref{gid}) for vector functions $\psi$, $\phi$.\\
State the uniqueness theorem. Specify what do you mean by uniqueness. \\
Hint: Find out a vector analog of the following scalar theorem used in electrostatics:\\
If $\nnabla \Phi = 0$, then $\Phi =$~const$(\br)$.
\\
{\sl References:  \cite{Kim}, Exercise~2.1 and Sec.~2.2.1, \cite{Poz}, Sec.~1.5.}

Compare your findings with results obtained by other groups.

Answer the guiding question.

\subsection{Boundary conditions}
{\sl In Stokesian hydrodynamics from now on we will restrict to the stick boundary conditions, i.e. to the fluid velocity $\bv(\br)$ at the boundary equal to the rigid motion velocity of the boundary itself:}
\be
\bv(\br) = \bU + \oOmega \times \br \label{rm}
\ee
\bq \footnotesize
\noi {\sl However, according to \cite{F76b}, \cite{F88}, the formalism is valid for a more general class of the so-called slip boundary conditions, namely
\bee
\btan \cdot \bv(\br) &=& \btan \cdot (\bU + \oOmega \times \br) + l \: \btan \cdot \ssigma(\br) \cdot \bn \nonumber\\
\bn \cdot \bv(\br) &=& \bn \cdot \bU  \hspace{6cm}\mbox{at the boundary S}\nonumber
\eee
where $\btan$ is a unit vector tangential to the boundary surface, $l$ vary from $0$ (for the stick boundary conditions) to $\infty$ (the so-called perfect slip boundary conditions), $\ssigma$ is the fluid stress tensor: $\sigma_{ij}=\mu (\partial_i v_j + \partial_j v_i) - p \delta_{ij}$.}

\paragraph*{A supplementary problem: slip boundary conditions}

{\sl In Stokesian hydrodynamics the slip boundary conditions on the sphere surface are defined as \cite{F76b}:
\bee
\btan \cdot \bv &-& \btan \cdot (\bU + \oOmega \times \br) = l \: \btan \cdot \ssigma \cdot \bn \\
\bn \cdot \bv &=& \bn \cdot \bU
\eee
where $\btan$ is a unit vector tangential to the boundary surface, $l$ vary from $0$ (the so-called no slip or stick boundary conditions) to $\infty$ (the so-called perfect slip boundary conditions).
}

Do slip conditions determine a unique solution to the Stokes equations? Prove your statement.
\eq

{\sl In Sec.~3 we assume that the fluid is motionless at infinity: $\bv|_{\infty}=0$. In the following sections we will consider the general 
case of any conditions at infinity.}

Construct simple electrostatic analogues of:
\\
A) ($\bv_i,\oOmega_i$) - a pair consisting of translational and rotational velocities of a body $i$,\\
B) a body at rest in a viscous fluid,\\
C) a rigidly moving body in a viscous fluid,\\
Specify analogous equations for the boundary conditions in all cases.

\subsection{Friction problem}
\noi {\sl The friction problem for N bodies is the following:\\
If given: $\oOmega_{\alpha}$, $\bU_{\alpha}$, $\bv _0(\br)$, then what are $\bF_{\beta}$, $\bT_{\beta}$ ($\alpha,\beta=1,...,N$)? \\
In this section we assume that the  ambient fluid flow vanishes: $\bv _0(\br)=0$. 
In such a case motion of N bodies with velocities 
${\bf U}_{\beta} + \oOmega_{\beta} \times ({\bf r}-{\bf r_{\beta}})$ result in  
forces ${\bf F_{\alpha}}$ and torques ${\bf T_{\alpha}}$ exerted on the fluid by 
the body ${\alpha}$, determined by the 
N-particle friction matrix $\xb$: 
\bee
\left( \ba{c}
{\bf F_{\alpha}}\\
{\bf T_{\alpha}} \ea \right) = \xb_{\alpha \beta} \left( \ba{c} {\bf U}_{\beta}\\ \oOmega_{\beta} \ea \right) \label{fm}
\eee
}

Explain how does (\ref{fm}) follow from Stokes equations. Does $\xb$ depend on: \\
- position in the fluid, ${\bf r}$, \\
- position of a body ${\beta}$, ${\bf r_{\beta}}$, \\
- translational velocity of a body $\beta$, ${\bf U}_{\beta}$ and \\
- rotational velocity of a body $\beta$, $\oOmega_{\beta}$? \\
If yes, explain how. If no, why not? What is the dimension of $\xb$?\\
Are the $\alpha \beta$ components of the 2-particle friction matrix equal to $\xb_{\alpha \beta}$, 
the corresponding components of the N-particle friction matrix from eq.~(\ref{fm})? Support your answer by a reasoning.

Develop an electrostatic analogue of the friction problem (make it as simple as possible). 
What are the electrostatic analogues of the quantities appearing in eq.~(\ref{fm})? Explain.

\noi {\sl In Sec. \ref{s3}-\ref{s4} we consider N rigid bodies of an arbitrary  shape. Later we will concentrate on N spheres only.}


\subsection{Green function}\label{s3.5}
Green function $G$ will help us to solve our friction problem, in a similar way it helps to solve its electrostatic analogue.
\subsubsection*{Guiding question}
Predict if the following statement is always true, true only under special supplementary conditions (if yes, specify them) or false;  explain your reasoning:
\be
\Psi(\br) = \int d^3\br' G({\bf r},{\bf r}') s({\bf r}')\label{gbc}
\ee
\subsubsection{Definition}
{\sl Green function $G$ is a solution to the equation:}
\be
L_0({\bf r})\; G({\bf r},{\bf r}') = \delta({\bf r}-{\bf r}')\label{defg}
\ee

Is $G$ a scalar, a vector or a tensor in: A) electrostatics; B) hydrodynamics?\\
Write down (\ref{defg}) explicitely for electrostatics and for hydrodynamics, indicating arguments and all components.

Specify $G$ in electrostatics (\cite{Jackson}, Sec.~1.10) and in hydrodynamics (\cite{Kim}, Sec.~2.4.1) if the Dirichlet boundary conditions vanish at infinity. Make your definition consistent with your choice of $L_0$. 

{\sl From now on we will assume that $G$ is the Green function for an infinite 
system.}
\bq \footnotesize
{\sl \noi However, the formalism presented in Sec. \ref{s3}-\ref{s4} has been developed for any Green function $G$ \cite{F88} -- corresponding to a container or to periodic boundary conditions \cite{Has}, \cite{F89}, \cite{CF89b}.
}
\eq

\subsubsection{Derivatives of the Green functions $\bG$ and $\bP$}
$\,$\indent {\sl The Green functions $\bG$ and $\bP$ satisfy the following identities:
\bee
\mu \nabla^2 {\cal G}_{ij} (\bR) - \partial_i{\cal P}_{j} (\bR) &=& - \delta_{ij}\delta^3(\bR)\label{d1}\\
\partial_i {\cal G}_{ij}(\bR) &=& 0\label{d2}\\
\nabla^2 {\cal P}_{j} (\bR) &=& \partial_j\delta^3(\bR)\label{d3}\\
\partial_j {\cal P}_{j} (\bR) &=& \delta^3(\bR).\label{d4}
\eee
where $\bR = \br' - \br$ and all derivatives are taken with respect to $\br'$: $\partial_i \equiv \partial/\partial r_i'$.

Although $\bG$ and $\bP$ are functions, but their derivatives are distributions.  We need to have a clear prescription how to evaluate such derivatives. Each of them can be understood as a limit of a sequence of functions.} 

Show that the Green functions $\bG$ and $\bP$ for the infinite system (our choice from Sec.~\ref{s3.5}):
\bee
{\cal G}_{ij} (\bR) &=& {1\over{8\pi \mu}} \left( {\delta_{ij}\over R} + {R_i R_j\over {R^3}} \right)\label{oseen}\\
{\cal P}_j (\bR) &=& {1\over{4\pi }} {R_j\over {R^3}} + {\cal P}_{0j}\label{pe},
\eee 
can be obtained as the following limits:
$\bG=\lim_{a\rightarrow 0} \bG^a$ and  $\bP=\lim_{a\rightarrow 0} \bP^a$, where 
\bee
{\cal G}^a_{ij}(\bR) &=&  {1\over{8\pi \mu}}(-\partial_i\partial_j + \delta_{ij}\nabla^2)\: (R^2+a^2)^{1/2}\\
{\cal P}^a_j(\bR) &=& -  {1\over{8\pi}} \partial_j \nabla^2\: (R^2+a^2)^{1/2}
\eee
Explain the procedure how to evaluate $D{\bG}$ and $D{\bP}$ for a differential 
operator $D$. \\
Apply this prescription to verify eq.~(\ref{d2}).

\subsection{Boundary integral equations}
{\sl Boundary integral equations are useful if there is a closed surface $S$ surrounding a volume $V$ and the field $\Psi$ is defined both inside and outside.}
\subsubsection{Electrostatics}

Use the Green's theorem:
\be
\int_V (\phi \nabla^2 \psi -  \psi \nabla^2 \phi)d^3 \br = \oint_S 
\left[\phi {\partial \psi\over {\partial n}} - \psi {\partial \phi\over {\partial n}}\right] dA\label{Gree}
\ee
to derive the following expressions for the electrostatic field potential $\Phi$ 
(\cite{Jackson}, Sec.~1.10):
\bee
\int_V \rho(\br')G (\br,\br') d^3 \br' + {1\over{4\pi}}\oint_S \left[G(\br,\br') 
{\partial \Phi(\br')\over {\partial n'}} - \Phi(\br') 
{\partial G(\br,\br')\over {\partial n'}}\right] dA' = \left\lbrace \ba{c} \Phi(\br) \hspace{0.3cm}\br \in V\\ 0 \hspace{1cm}\br \notin V\ea\right.\nonumber\\
\label{zero}
\eee

\noi {\sl Reference: \cite{Jackson}, Sec.~1.6, 1.8.}

What is the hydrodynamic analogue of the electrostatic ${\partial \Phi(\br)\over {\partial n}}$?

Does (\ref{zero}) answer the guiding question? Explain.
\subsubsection{Stokesian hydrodynamics}\label{s3.6.2}

{\sl If there is no ambient flow, the hydrodynamic analogue of identities (\ref{zero}) is given as:}

\bee
\int_V \rho f_j {\cal G}_{jk} d^3 \br' + \oint_S \{{\cal G}_{jk} \sigma_{lj} - v_j [\mu (\partial'_l {\cal G}_{jk} + \partial'_j {\cal G}_{lk}) - \delta_{jl} {\cal P}_k]\} n'_l dA' &=& \left\lbrace 
\ba{l} 
v_k(\br) \hspace{1.35cm } \br \in V\\ 
0  \hspace{2cm}\br \notin V
\ea \right. \nonumber \\
\label{Gi}\\
-\int_V \rho f_j {\cal P}_j d^3 \br' + \oint_S \{-{\cal P}_j \sigma_{lj} + 
\mu v_j ((\partial'_l {\cal P}_{j} + \partial'_j {\cal P}_{l})] \} n'_l dA' &=& \left\lbrace \ba{l} p(\br) \hspace{1.3cm } \br \in V\\
0 \hspace{1.8cm } \br \notin V \ea \right.\nonumber \\
\label{Pi}
\eee

Are $\ssigma$ and $\bv$ taken at $\br$ or $\br'$?
Put the missing order of arguments of the Green functions $\bG$ and $\bP$ 
-- ($\br$, $\br'$) or ($\br'$, $\br$) -- into eqs~(\ref{Gi})-(\ref{Pi}). 

What are the symmetry properties of the Green functions ${\bG}$ and ${\bP}$? 
Which of them are general, and which are due to the specific symmetries (no fluid motion at infinity) of the Oseen functions (\ref{oseen})-(\ref{pe})?

How does the unit vector $\bn'$ point: out or into the fluid?

How do the equations simplify if there is no external forces acting on the fluid other than gravity?

\noi {\sl Reference: \cite{Poz}, Sec.~2.3, \cite{Kim}, Sec.~2.4.2, \cite{HB}, Sec.~3.4.}

Eqs (\ref{Gi})-(\ref{Pi}) are valid if there is a closed boundary of any shape inside a fluid. How would you modify them to describe a rigid 
body in a fluid? Explain. 

{\sl The integral representation (\ref{Gi})-(\ref{Pi}) still does not allow to address the quiding question from Sec.~\ref{s3.5} -- just the opposite, it seems to contain a term differing in form from eq.~(\ref{gbc}). We will come back to this problem in Sec.~\ref{sn42}, but first we will learn how to take into account the existence of an ambient flow around a particle.}

\section{Particle in ambient flow}\label{s4}
\hfill{\footnotesize \it \parbox{7.5cm}{{\em Auguste Rodin}:  ``First, I usually create my stone children without cloths. Then all I have to do is to throw some drapery over them...\cite{Roda}
''}}

\vspace{0.5cm}
\subsection{Ambient flow}\label{sn41}
How to use the results of Sec.~\ref{s3} to construct solutions of the Stokes equation in the presence of an ambient flow?

\subsubsection{Definitions}\label{sn411}
{\sl 1. The ambient flow is a solution of the Stokes equation with given
boundary conditions.\\
2. The external ambient flow $\bv_0(\br)$ is a solution of the Stokes equation with a given
boundary condition at infinity: 
\be
\bv_0(\br)|_{\infty} = \bV(\theta, \phi)\label{ambient}
\ee}
Specify what is the ambient flow for the boundary condition:
\be
\bv_0(\br)|_{\infty} = \bV_0.\label{stala}
\ee
Explain your reasoning. \\
Give examples of other external ambient flows.\\
Construct example of an ambient flow which has the same boundary conditions at infinity as a certain external ambient flow but which is different. Predict  how this construction can be in agreement with the uniqueness theorem from Sec.~\ref{uniq}. ({\sl We will come back to this example in Sec.~\ref{sn421}}.) 

\subsubsection{Equvalence of solutions}\label{equiv}
Assume that $\bv(\br)$ is the solution of the Stokes equation with the 
boundary conditions at infinity and on closed surfaces $S_{\alpha}$,  $\alpha=1,...,N$:
\bee
\bv(\br)|_{S_{\alpha}}&=& \bU + \oOmega \times (\br|_{S_{\alpha}}-\br_{\alpha})\label{bc1}\\
\bv(\br)|_{\infty} &=& \bV(\theta, \phi)\label{bc2}
\eee
{\sl It is often said that the bodies $S_{\alpha}$ are immersed in the ambient flow $\bv_0(\br)$ given by (\ref{ambient}).}

Does $\bv(\br)-\bv_0(\br)$ satisfy the Stokes equation? If yes, then specify the boundary conditions and explain your reasoning. If no, then why not.
Explain how two solutions -- in the presence and in the absence of an ambient flow -- are equivalent.

\subsection{Formalism of induced forces}\label{sn42} 
Generalize the eqs (\ref{Gi})-(\ref{Pi}) from Sec.~\ref{s3.6.2} to describe a closed boundary  in  an ambient external flow $\bv_0(\br)$. Explain your reasoning.

{\sl The starting point of \cite{F88} (eq.~2.7) is the conjecture that the formal solution to the Stokes equation can be written as:
\be
\bv(\br) - \bv_0(\br) = \int \bG (\br,\br') \bcF(\br') d^3 \br'\label{F2.7}
\ee
where $\bcF(\br)$ is the total force density 
exerted on the fluid. }

{\sl To understand and to justify this statement we will derive a formula in the form of (\ref{F2.7}) from the boundary integral equations given in Sec.~\ref{s3.6.2}. To this goal we will first develop the concept of induced surface force density.}

\subsubsection{The concept of induced forces}\label{sn421}
Construct a hydrodynamic analogue of the electrostatic relation between the induced surface charge density and the electrostatic field.  How do you interpret the meaning of the adjective {\it induced} describing 
a surface density on the boundary in both cases? 

\noi {\sl The goal is to express the boundary integrals given  in Sec.~\ref{s3.6.2} in terms of a surface induced force density (depending on the fluid stress tensor at the surface) rather than in terms of the fluid velocity $\bv$ at the boundary.

The question is how to achieve it - the eqs (\ref{Gi})-(\ref{Pi}) depend on both $\ssigma$ and $\bv$ at the boundary. The idea is to first introduce an artificial fluid flow inside the rigid solid particles. Namely, to construct the inside solution to the Stokes 
equation 
with the same stick boundary conditions (\ref{bc1}) at the particle surface as those which determine the real outside solution. Next, to determine what are these induced forces, using the eqs (\ref{Gi})-(\ref{Pi}), which you have just generalized for a non-vanishing ambient flow. 

\noi Reference: \cite{CB}, Sec.~2 and \cite{MB}, Sec.~3-4.}

Solve the Stokes equation inside a volume $V$ with the stick boundary 
conditions 
(\ref{bc1}) and the ambient flow given by (\ref{bc2}). Hint: Make use of the uniqueness theorem.
{\small \paragraph*{A supplementary problem: uniqueness theorem revised.} The combination of the inside and the outside solutions is different than the ambient flow, although both satisfy the same boundary conditions at infinity. 
Explain how this construction can be in agreement with the uniqueness theorem from Sec.~\ref{uniq}. Compare with your reasoning from 
Sec.~\ref{sn411}, where you also constructed example of an ambient flow which 
had the same boundary conditions at infinity as a certain different external ambient flow. }

\subsubsection{Justification of the formalism}
Write down two sets of the generalized eqs (\ref{Gi})-(\ref{Pi}): \\
- for V being the interior of all the particles and \\
- for V being the real fluid.\\
Combine both sets and make use of the stick boundary conditions to eliminate the surface integrals including values of the fluid velocity at the boundary. \\
Derive an expression for the induced force density in terms of the fluid 
stress tensor at the surface. 

How will you calculate the total force acting on a body in terms of the fluid stress tensor, and how in terms of the induced force density? 

How does the induced force density relate to $\bcF$ in eq.~(\ref{F2.7})? 
What is the 
range of integration in eq.~(\ref{F2.7})? 

Use the generalized eqs (\ref{Gi})-(\ref{Pi}) to derive an equation for $p(\br)$ similar to  eq.~(\ref{F2.7}).  

Specify the properties of the Green function used in this section. How could they  be justified?\\

\subsection{Method of reflections}

\subsubsection{The difficulty}
Assume that $\mbox{\bf v}_{\alpha}$ is the solution of the one-particle friction problem, i.e. the solution to the Stokes equation with the boundary conditions (\ref{bc2}) and the one-particle version of (\ref{bc1}). 

Does $\mbox{\bf v} = \sum_{\alpha=1}^N \mbox{\bf v}_{\alpha}$ satisfy Stokes equations? 
Explain your reasoning.\\
Does $\mbox{\bf v}$ satisfy the boundary conditions (\ref{bc2}) at infinity? If 
yes, then why? If not, then how could you construct $\mbox{\bf v'}$, another
combination of  $\mbox{\bf v}_{\alpha}$, satisfying (\ref{bc2})?
\\
Calculate $\mbox{\bf v}$ (and $\mbox{\bf v'}$) at the surface $S_{\alpha}$ of 
the body $\alpha$. 
How do they compare to the boundary conditions (\ref{bc1})?\\
Are $\mbox{\bf v}$ (and $\mbox{\bf v'}$) 
solutions to the N-particle friction problem? Explain your reasoning. \\

\subsubsection{Construction}
{\sl Method of reflections is an iteration procedure to construct 
an approximate  solution to the N-particle friction problem, building it 
from N single-particle solutions.
At each step corrections are added to decrease descrepency between the boundary conditions and the actual value of the approximate solution on the body surfaces. It means that at each step we modify the single particle solutions 
by a better adjustement of their boundary conditions. 
The N-particle solution $\bv$ is formally written as:
\be
\bv = \sum_{\alpha=1}^N \bv_{\alpha}
\ee
Each single particle solution $\bv_{\alpha}$ for particle $\alpha$ is given by the following formal expansion:
\be
\bv_{\alpha} = \bv_0 + \bw_{\alpha} +  \sum_{\beta \neq \alpha} \bw_{\beta \alpha} +  \sum_{\gamma \neq \beta}\sum_{\beta \neq \alpha} \bw_{\gamma \beta \alpha} + ...\label{expan}
\ee
Each $\bw_{\gamma ... \alpha}$ satisfies the Stokes equations.}\\
{\sl Reference: \cite{Kim}, Sec.~8.1.
}

Specify what are the boundary conditions at infinity and at the surface 
$S_{\alpha}$ for $\bv_0$, for $\bw_{\alpha}$, for $\bw_{\beta \alpha}$ and 
for 
$\bw_{...\beta \alpha}$. 

What is the approximate value of $\bv$ at the surface $S_{\alpha}$ after: \\
- the first\\
- the second\\
- the n-th\\
iteration step? 

What is the effective ambient flow in which the particle $\alpha$ is immersed 
before:\\ 
- the first, \\
- the second and \\
- the n-th \\
iteration step? 

\subsubsection{Interpretation}\label{sn433}
{\sl Eq. (\ref{expan}) can be interpreted as a summation over all incident and outgoing "waves" in multiple scattering (in a sequence of reflections).}

Specify what are the incident and the outgoing "waves" scattered by a particle $\beta$ at the first, the second and the n-th iteration step. 

Predict if the iteration procedure is convergent. Give arguments.

\part{Exploring the structure: a lecture.}\label{les}
\hfill{\footnotesize \it \parbox{7cm}{{\em Auguste Rodin:} ``For the first 
time I saw separate pieces, arms, heads or feet; then I attempted the 
figure as a whole. Suddenly, I grasped what unity was...''\cite{Rodf}}} 

\vspace{0.5cm}
\section{Application of the basic concepts}

Reference: \cite{F88}, Sec.~2.

\subsection{Reformulation of the friction problem}\label{sn51}
Forces and torques exerted by the fluid on the surface $S_{\alpha}$ of the particle $\alpha$  are given in terms of the fluid velocity $\bv$ and pressure $p$ as:
\bee
{\bf F_{\alpha}} &=& - \oint_{S_{\alpha}} \ssigma \cdot {\bf n_{\alpha}} \: dA \label{sila}\\
{\bf T_{\alpha}} &=& - \oint_{S_{\alpha}} ({\bf r}-{\bf r}_{\alpha}) \times (\ssigma \cdot {\bf n_{\alpha}}) \: dA\label{moment}
\eee
where $\bv$, $p$ satisfy the Stokes equations: 
\bee
\mu \nabla^2\bv-\nnabla p &=& 0 \label{Stokes} \\
\nnabla \cdot \bv &=& 0 \label{incomp}
\eee
with given boundary conditions:
\bee
\bv(\br) &\rightarrow&  \bV(\theta, \phi)\hspace{6cm}r\rightarrow \infty\label{inf}\\
\bv(\br) &=& \bw_{\alpha}(\br) \equiv \bU_{\alpha} + \oOmega_{\alpha} \times (\br - \br_{\alpha})\hspace{2cm}\br \in S_{\alpha}\label{par}
\eee
The surface normal $\bn$ in (\ref{sila})-(\ref{moment}) points into the particle.

To solve (\ref{sila})-(\ref{par}) we will apply the tools developed so far, in the following way:\\
1. We evalute the ambient flow ($\bv_0$, $p_0$) as the solution to (\ref{Stokes})-(\ref{incomp}) 
with (\ref{inf}).\\
2. We use the formalism of induced forces (Sec.~\ref{sn42}) and the concept of equivalent solution (Sec.~\ref{sn41}) to replace (\ref{sila})-(\ref{par}) by: 
\bee
{\bf F_{\alpha}} &=& \int {\bcF} \:\, d^3\br \label{sila2}\\
{\bf T_{\alpha}} &=& \int ({\bf r}-{\bf r}_{\alpha}) \times {\bcF} 
\: d^3\br\label{moment2}\\
\bw_{\alpha}(\br) - \bv_0(\br) &=& \int \bG (\br,\br') \bcF(\br') d^3 \br' 
\hspace{2cm} \mbox{for } |\br-\br_{\alpha}| \le a 
\label{mai}
\eee
with $\bG$ given by (\ref{oseen}), and with $\bcF$ -- non-vanishing on the particle surfaces only -- related to the fluid stress tensor $\ssigma$ as in \cite{F76a}:
\bee
\bcF(\br) &=& \sum_{\alpha}\bcF_{\alpha}(\br) \label{sumf}\\
\bcF_{\alpha}(\br) &=&  \ssigma \cdot {\bf n}_{\alpha} \: \delta(|\br-\br_{\alpha}|-a)
\eee
Eq. (\ref{mai}) has beed obtained by substitution of (\ref{par}) into (\ref{F2.7}). 

\subsection{Induced forces in terms of the boundary conditions for the fluid velocities}
The reformulated friction problem means solving (\ref{mai}) for $\bcF$. 
First we note that due to linearity of the Stokes equations $\bcF$ depends linearly on the boundary conditions of the equivalent solution, i.e. the rigid motion of the spheres minus the ambient flow: $\bw(\br) - \bv_0(\br)$, where
\be
\bw(\br) = \sum_{\alpha} \bw_{\alpha}(\br) \label{sumw}
\ee
and $\bw_{\alpha}$ are given by (\ref{par}).
We write it as:
\be
\bcF(\br) =\int \bZ (\br,\br') [\bw(\br') - \bv_0(\br')  ] d^3 \br'\label{mistry}
\ee
and $\bZ$ vanishes if  $\br$ or $\br'$ is located outside a particle. 

$\bZ$ has a matrix form, with $\bZ_{\alpha \beta}$ relating $\bcF_{\alpha}$, the forces acting on the particle $\alpha$, to the boundary conditions $\bw_{\beta} - \bv_0$, on the surfaces of all the particles $\beta$. Using a simplified notation we write it as:
\be
\bcF_{\alpha}  =\bZ_{\alpha \beta}  [\bw_{\beta}  - \bv_0]\label{duzeZ}
\ee
Therefore our goal is to find the N-body friction kernels $\bZ_{\alpha \beta}(\br, \br')$. We will do it in two steps: 

1. Simplification: one particle $\alpha$ in the fluid flow. We will have a look how to evalute the one-particle friction kernel $\bZ_0(\alpha)$. 

2. Multiplication: many particles. We will see how to express $\bZ$ in terms of one-particle friction kernels $\bZ_0(\alpha)$.

\subsection{Multipole expansion}\label{sn53}
Assume that we know the N-particle friction kernel $\bZ_{\alpha \beta}$. The question is how to find the forces and the torques. Substituting  
(\ref{mistry}) into (\ref{sila2}) and (\ref{moment2}) we get:
\bee
{\bf F_{\alpha}} &=& \int d^3 \br \int d^3 \br' \bZ_{\alpha \beta} (\br,\br') [\bw_{\beta}(\br') - \bv_0(\br')] \label{fa}\\
{\bf T_{\alpha}} &=& \int d^3 \br \int d^3 \br' (\br - \br_{\alpha}) \times \bZ_{\alpha \beta} (\br,\br') [\bw_{\beta}(\br') - \bv_0(\br')] \label{ta}
\eee

If there is no ambient flow, then:
\be
\bw_{\alpha}(\br) - \bv_0(\br) = \bU_{\alpha} + \oOmega_{\alpha} \times (\br - \br_{\alpha})
\ee
and we have:
\bee
\left( \ba{c}
{\bf F_{\alpha}}\\
{\bf T_{\alpha}} \ea \right) =\left(\ba{cc} \xb^{tt}&\xb^{tr}\\ \xb^{rt}&\xb^{rr}\ea \right)_{\alpha \beta} \left( \ba{c} {\bf U}_{\beta}\\ \oOmega_{\beta} \ea \right) \label{fmo}
\eee
The friction matrix elements can be written as:
\be
<\bb_{i\alpha}|\bZ_{\alpha \beta}|\bb_{j\beta}>
\ee
with
\bee
<\bb_{0\alpha}| = |\bb_{0\alpha}>  &=& \bone \: \theta_{\alpha}(\br) \\
<\bb_{1\alpha}| = |\bb_{1\alpha}> &=& - \epsilon_{ijk}(\br - \br_{\alpha})_k \theta_{\alpha}(\br) 
\eee
and the scalar product defined as:
\be
<\baa\;|\;\bb> = \int \baa^*(\br) \cdot \bb(\br) d^3\br
\ee

If there is no ambient flow, then:
\be
\bw_{\beta}(\br) - \bv_0(\br) = \sum {(\br -\br_{\beta})^p\over{p!}}\nnabla^p[\bw_{\beta}(\br_{\beta}) - \bv_0(\br_{\beta})]
\ee
We construct a complete set of functions $\bb_{\beta}^p$ -- combinations of $(\br -\br_{\beta})^p$ and a complete set of "velocity multipoles" $\bC_{\beta}^p$ -- combinations of $\nnabla^p[\bw_{\beta}(\br_{\beta}) - \bv_0(\br_{\beta})]$. Instead of eq.~(\ref{fmo}) we now have:
\bee
\bF_{\alpha} &=& <\bb_{0\alpha}|\bZ_{\alpha \beta}|\bb_{p\beta}>\bC_{\beta}^p\\
\bT_{\alpha}  &=& <\bb_{1\alpha}|\bZ_{\alpha \beta}|\bb_{p\beta}>\bC_{\beta}^p
\eee
or equvalently:
\bee
\left( \ba{c}
{\bf F_{\alpha}}\\
{\bf T_{\alpha}} \ea \right) =\left(\ba{ccc} \xb^{tt}&\xb^{tr}&...\\ \xb^{rt}&\xb^{rr}&...\ea \right)_{\alpha \beta}   \bC_{\beta}^p \label{fmo2}
\eee
In particular:
\bee
\bC_{\beta}^0 = \bU_{\beta}\\
\bC_{\beta}^1 = \oOmega_{\beta} + {1\over 2} \nnabla \times \bv_0
\eee
The questions remain how to choose the basis $\bb_{\beta}^p$, how to expand the N-particle friction kernel $\bZ$ and how to truncate. First we need to get acquinted with the N-particle and the one-particle friction kernels $\bZ$ and $\bZ_0$.

\subsection{Single particle solution}
The single particle solution in an ambient flow $\bw_0$ is given as (\ref{F2.7}), (\ref{mistry}):
\be
\bv(\br) - \bw_0(\br) = \int \left[ \int \bG (\br,\br'') 
\bZ_0 (\br'',\br') d^3 \br''\right] [\bw(\br') - \bw_0(\br')] d^3 \br'
\label{bb}
\ee
Note that $\br',\br''$ are inside the particle, while $\br$ has no such restriction. 
To shorten notation we write (\ref{mistry}) and (\ref{bb}) as:
\bee
\bcF  &=&\bZ_{0}  [\bw  - \bw_0]\label{shf}\\
\bv - \bw_0 &=& \bG  
\bZ_{0}  [\bw - \bw_0] \label{shv}
\eee

\subsection{Multiple scattering}
We can describe our many-particle system as the system consisting of a sphere 
$\alpha$ in an ambient flow created by the other bodies.
But this 
ambient flow is also unknown, and it depends on the position and the 
motion of the sphere $\alpha$.
To solve this problem we will construct an iteration procedure. At each step the velocity field will be approximated by the sum of all one-particle solutions $\bv_{\alpha}$ in a given ambient flow, evaluated from (\ref{shv}). Since this is not a many-particle solution, then $\bv_{\beta}$ will change the ambient flow in which particle $\alpha$ is immersed. It will be taken into account through modification of the ambient flow 
entering the next step of the iteration.

The interpretation of this iteration in terms of a 
"multiple 
scattering" (or "reflection") on each particle $\beta$ has been made 
in Sec.~\ref{sn433}. 

To carry out the multiple scattering we need to specify what is the total ambient flow $\bw_n$, in which the particle $\alpha$ is immersed after each step $n$ of the iteration 
procedure. $\bw_n$ is the "wave" outgoing from the step $n$ and incident to step $n+1$.
It consists of the incident ambient flow $\bw_{n-1}$ and corrections coming 
from the other particles $\beta \neq \alpha$, evaluated from 
eq.~(\ref{shv}) in step $n-1$.  
Therefore: 
\be
\bw_{n} = \sum_{\beta \neq \alpha} [\bv_{n \beta}-\bw_{n-1}] + \bw_{n-1}\label{care}
\ee
$\bv_{n \alpha}-\bw_{n-1}$, $n=1,2,...$ correspond to the subsequent terms in the scattering expansion given in eq.~(\ref{expan}). In Sec.~\ref{sn433} we have already analyzed the boundary conditions for them. Therefore,  
with the help of (\ref{shv})-(\ref{care}) the multiple scattering process made in step $n$ can be described as: 
\bee
\bv_{1\alpha} - \bv_{0} &=&  \bG  
\bZ_{0\alpha}  [\bw_{\alpha} - \bv_{0}] \label{nowykrok1}\\
\bv_{(n+1)\alpha} - \bw_{n} &=&  - \bG  
\bZ_{0\alpha}  (\bw_{n}-\bw_{n-1}) \hspace{2cm}\mbox{for }\; n \ge 2
\label{nowykrok}
\eee
The multiple scattering (\ref{care})-(\ref{nowykrok}) is equivalent to the construction of the fluid velocity $\bv$ by 
the metod of reflections \cite{Kim}, \cite{HB}. Eqs (\ref{care}),(\ref{nowykrok1}) and (\ref{nowykrok}) specify the ambient flow:
\bee
\bw_{1} - \bv_0 &=& \sum_{\beta \neq \alpha} \bG  
\bZ_{0\beta}  [\bw_{\beta} - \bv_0]\label{af1}\\
\bw_{n+1} - \bw_{n} &=& - \sum_{\beta \neq \alpha} \bG  
\bZ_{0\beta}   (\bw_{n}-\bw_{n-1})  \hspace{2cm}\mbox{for }\; n \ge 2
\label{af}
\eee
and eq.~(\ref{shf}) gives the induced forces:
\bee
\bcF_{1\alpha} &=& \bZ_{0\alpha} [\bw_{\alpha} - \bv_{0}]\\
\bcF_{(n+1)\alpha}  &=& \bZ_{0\alpha}   (\bw_{\alpha}-\bw_{n}) = 
\bcF_{n\alpha}  + \bZ_{0\alpha}   (\bw_{n-1}-\bw_{n})
\hspace{2cm}\mbox{for }\; n \ge 2
\eee
Therefore the multiple scattering expansion of $\bcF_{\alpha}$ reads to the following form of the N-particle friction kernel $\bZ_{\alpha\beta}$ in eq.~(\ref{duzeZ}):
\bee
\bZ_{\alpha\beta} &=& \bZ_{0\alpha} \delta_{\alpha\beta} - (1-\delta_{\alpha\beta}) \bZ_{0\alpha}  \bG \bZ_{0\beta}  + \bZ_{0\alpha} \sum_{\gamma \neq \alpha, \beta} \bG \bZ_{0\gamma}  \bG \bZ_{0\beta} - ...
\eee

\hfill{\footnotesize \it \parbox{7.5cm}{{\em Auguste Rodin}:  ``Yes, form I have looked at and understood, it can be learnt: but the genius of form has yet to be studied. 
''\cite{Rodf}}}

\section*{Acknowledgements} 

I thank Fran\c cois Feuillebois for inviting me to guide the course on hydrodynamic interactions between many spheres -- and the same for providing me the motivation to develop the work presented here. 
I benefited from the educational structure of  Physique Thermique, Laboratoire de 
Physique et Mecanique des Milieux Heterogenes, \'Ecole Sup\'erieure de Physique et de Chimie Industrielles de la Ville de  Paris, where 
the sessions took place 
in the framework of a non-standard educational activity.  
Discussions with 
the participants of the course, Paul Chaikin, Fran\c cois Feuillebois, Pierre-Emmanuel Jabin, 
Nicolas Lecoq, Daniel Lhuillier, Michel Martin were essential for developing this paper, although I remain responsible for all the possible mistakes. 
I thank Stanis\l aw G\l azek, Tomasz Mas\l owski and Marek Wi\c eckowski for sharing with me their teaching experience. My stay at ESPCI has been supported by the French Ministry of Education and Research.
\\

\section*{Appendix. Announcement about the course}
{\small Between June 15 and July 10, 1998 I will guide a course on\\

HYDRODYNAMIC INTERACTIONS BETWEEN MANY SPHERES\\

4 sessions, 2.5 hours each.\\

GOAL: \\
To inquire the basic structure and tools of the modern theory, which 
has been developed by Felderhof, Jones, Cichocki, Schmitz and their 
coworkers for 20 years, and resulted in numerical packages allowing
for accurate calculations of hydrodynamic interactions.\\

SUBJECT: \\
How to determine the behavior of N spheres in low Reynolds number 
incompressible flow (N between several and several hundred). Namely: \\
If an ambient fluid flow and external forces \& torques acting on them 
are given, then what are their translational \& rotational velocities 
(mobility problem)?\\
If their translational \& rotational velocities and an ambient fluid 
flow are given, then what are the forces \& torques they exert on the 
fluid (friction problem)?\\

KEY WORDS: \\
Stokes equations, stick boundary conditions, ambient flow, Green 
function, induced forces, friction kernel, generalized resistance matrix, 
generalized mobility matrix, multiple scattering expansion, force 
multipole moments, vector harmonics, rotational invariance.  \\

IDEA:\\
To simplify, but knowing how to reach for complexity. \\
Since the course is meant to be a first step needed to be done before 
making a more sophisticated analysis, then we will concentrate on basic 
concepts applied to a simple system. In particular, the following 
problems treated by this theory will be mentioned, but will not be 
discussed: slip boundary conditions, Green function other than Oseen 
tensor, lubrication phenomena, averaging procedure leading to evaluation 
of transport coefficients, mobility and friction problem for non-spherical 
shapes of particles (built from spheres). 
Analogy with electrostatics and quantum mechanics will be outlined, since 
you may later find it helpful in carring out calculations. \\

ATTITUDE:\\
To make it useful. \\
Therefore first of all you are welcome to participate in making a plan of 
the course, by e-mailing me your suggestions what you would like to gain, 
what do you need it for and which concepts from those listed above are of 
your interest and which are not. 
Secondly, to help you in applying the technique, the sessions will be 
based on your active inquiry in small groups and your own solving of some 
basic problems rather than on passive listening to a lecture. 
Finally, your comments on time allocated to this activity are appreciated.\\

CONTACT:\\
If you want to participate, reply by e-mail before Thursday, June 11.
Please let me know what are your time limitations - it will help me to 
fix the day of the week and the hour of our sessions. \\

Maria Ekiel-Je\.zewska}

\end{document}